\documentclass[aps,preprint,floatfix,nofootinbib,showpacs]{revtex4-1}

\pdfoutput=1
\usepackage{graphicx,color,float}
\usepackage{hyperref}
\usepackage{multirow}
\usepackage{verbatim}
\usepackage{longtable}
 \usepackage{amsmath}
\usepackage{amsfonts}
\usepackage{amssymb}
\usepackage{rotating}

\newcommand{\lsim}{\raisebox{-0.13cm}{~\shortstack{$<$ \\[-0.07cm] $\sim$}}~}
\newcommand{\gsim}{\raisebox{-0.13cm}{~\shortstack{$>$ \\[-0.07cm] $\sim$}}~}

\begin{document}

{\small
\begin{flushright}
IUEP-HEP-19-02
\end{flushright} }

\title{
Probing Trilinear Higgs Self--coupling at the HL-LHC\\ via Multivariate Analysis
}

\def\slash#1{#1\!\!/}

\renewcommand{\thefootnote}{\arabic{footnote}}

\author{
Jung Chang$^{1}$, Kingman Cheung$^{2,3,4}$, Jae Sik Lee$^{1,5}$, 
and Jubin Park$^{5,1}$}
\affiliation{
$^1$ Department of Physics, Chonnam National University, \\
300 Yongbong-dong, Buk-gu, Gwangju, 500-757, Republic of Korea \\
$^2$ Physics Division, National Center for Theoretical Sciences,
Hsinchu, Taiwan \\
$^3$ Division of Quantum Phases and Devices, School of Physics, 
Konkuk University, Seoul 143-701, Republic of Korea \\
$^4$ Department of Physics, National Tsing Hua University,
Hsinchu 300, Taiwan \\
$^5$ Institute for Universe and Elementary Particles, Chonnam National University, \\
300 Yongbong-dong, Buk-gu, Gwangju, 500-757, Republic of Korea
}
\date{October 30, 2019}

\begin{abstract}
We perform a multivariate analysis of Higgs-pair production 
in $HH \to b\bar b \gamma\gamma$ channel at the HL-LHC to probe
the trilinear Higgs self--coupling $\lambda_{3H}$, which takes
the value of 1 in the SM.
We consider all the known background processes. Also, for the signal we 
are the first to adopt the most recent event generator 
of {\tt POWHEG-BOX-V2} to exploit the NLO distributions 
for Toolkit for Multivariate Data Analysis (TMVA),
taking account of the full top--quark mass dependence.
Through Boosted Decision Tree (BDT) analysis trained for $\lambda_{3H}=1$,
we find that the significance can reach up to 1.95 with about $9$ signal and $18$
background events.
In addition, the Higgs boson self-coupling can be constrained to 
$1.00 < \lambda_{3H} < 6.22$
at 95\% confidence level (CL).
We also perform a likelihood fitting of $M_{\gamma\gamma bb}$ distribution 
and
find the $1\sigma$ confidence interval (CI) of
$0.1 < \lambda_{3H} < 2.2\, \cup\,5.4 < \lambda_{3H} < 6.6$ 
for the $\lambda_{3H}=1$ nominal set.
On the other hand, using BDTs trained for each value of $\lambda_{3H}$, 
we find a bulk region 
of $0.5 \lsim \lambda_{3H} \lsim 4.5$, for which it is hard 
to  pin down the trilinear coupling.

\end{abstract}

\maketitle

\section{Introduction}
Since the discovery of the Higgs boson in 2012 \cite{discovery},
the most pressing question is to
understand the underlying mechanism for electroweak symmetry breaking (EWSB).
There is no particular reason why the EWSB sector only consists of a single Higgs doublet.
Indeed, the simplest version suffers from the so-called gauge hierarchy problem.
After completing Run I and II at the LHC, the identity of the Higgs boson has been established.
It is best described as the standard model (SM) Higgs boson \cite{higgcision},
although there is an upward trend in the overall signal strength \cite{Cheung:2018ave}.

All current measurements of the Higgs boson properties confine to the couplings
of the Higgs boson to the SM particles, like gauge bosons and fermions .
However,  the self-couplings of the Higgs boson is not established at all,
which depends on the dynamics of the EWSB sector. 
The self-couplings of the Higgs boson can be very different between the SM and other
extensions of the EWSB sector, like two-Higgs doublet models (2HDM), and MSSM.
Higgs boson pair production at the LHC provides a very useful avenue to
investigate the self-couplings of the Higgs boson \cite{hh-early,hh-mid,hh-later}.
There have been a large number of works in literature on Higgs-pair production in the
SM \cite{hh-sm}, in model-independent formalism \cite{hh-in}, in models beyond the
SM \cite{hh-bsm}, and in SUSY \cite{hh-susy}.

Furthermore, the high luminosity option of the LHC running at 14 TeV (HL-LHC) was approved.
It is a legitimate machine to investigate the EWSB sector. 
In a previous work \cite{Chang:2018uwu}, we showed that even with the HL-LHC one
cannot establish the self-coupling $\lambda_{3H}$ at the SM value using the most
promising decay mode $HH \to (b\bar b)(\gamma\gamma)$.  Indeed, one can only constrain
the self-coupling to be within $-1.0 \alt \lambda_{3H} \alt 7.6$ at 95\% CL
\cite{Chang:2018uwu}.  These results were based on a conventional cut-based analysis.

In this work, we show that with the use of Boosted Decision Tree (BDT) method, the
significance of the signal can be improved by 80\%, which is a substantial improvement
from the cut-based analysis.

The organization is as follows. In the next section, we give some details on generation
of the signal and background event samples. In Sec. III, we set up the TMVA variables
and various BDT methods.  We present the numerical results in Sec. IV.  We end
our discussion and conclusion in Sec. V.

\section{generation and simulation of signal and backgrounds}
%
\begin{table}[t!]
\caption{Monte Carlo samples used in Higgs-pair production
  analysis $H(\rightarrow b\bar{b})H(\rightarrow \gamma\gamma)$, and
  the corresponding codes for the matrix-element generation, parton
  showering, and hadronization. The third (fourth) column shows their cross
  section times branching ratio (the order in perturbative QCD of the
  cross section calculation applied), and the final column shows their
  PDF set used in the simulation.
}\vspace{3mm}
\label{tab:ParticleList}
\centering
\begin{tabular}{  c  c  c  c  c  c }
\hline\hline 
\multicolumn{6}{c}{Signal} \\
\hline
\multicolumn{2}{c}{Signal process} & Generator/Parton Shower &
$\sigma \cdot BR$ [fb] & Order  & PDF used  \\
&&&& in QCD &
\\
\hline
\multicolumn{2}{c}{$gg \to HH \to b b \gamma\gamma$ 
} & $\mathtt{POWHEG}$-$\mathtt{BOX}$-$\mathtt{V}2$/$\mathtt{PYTHIA8}$  & 0.096 & NNLO & 
{\tt PDF4LHC15$\_$nlo} \\ 
\hline
\hline
\multicolumn{6}{c}{Backgrounds} \\
\hline
Background(BG)  & Process  & Generator/Parton Shower & $\sigma\cdot BR$~[fb] &  Order  &
PDF used\\
&&&& in QCD &   \\
\hline
\multirow{4}{*}{ }
& $ggH(\rightarrow \gamma\gamma)$  &  $\mathtt{POWHEG}$-$\mathtt{BOX}$/$\mathtt{PYTHIA6}$ & $1.20 \times 10^2$ & NNNLO & $\mathtt{CT10}$ \\ \cline{2-5}
Single-Higgs  & $t \bar{t} H(\rightarrow \gamma\gamma)$ & $\mathtt{PYTHIA8}$/$\mathtt{PYTHIA8}$  & 1.37 & NLO & \\ \cline{2-5}
   associated BG 
     &  $ZH(\rightarrow \gamma\gamma)$   &  $\mathtt{PYTHIA8}$/$\mathtt{PYTHIA8}$   & 2.24 & NLO &\\ \cline{2-5}
     & $b\bar{b}H(\rightarrow \gamma\gamma)$ &  $\mathtt{PYTHIA8}$/$\mathtt{PYTHIA8}$ & 1.26 & NLO  &\\ \hline
\multirow{7}{*}{Non-resonant BG} & $b\bar{b} \gamma\gamma$ & $\mathtt{MG5\_aMC@NLO}$/$\mathtt{PYTHIA8}$ &
$82.52$   &  LO & ${\mathtt{CT14LO}}$ \\ \cline{2-5}
                  &  $c\bar{c} \gamma\gamma$ & $\mathtt{MG5\_aMC@NLO}$/$\mathtt{PYTHIA8}$ & 
$647.3$ & LO &  \\ \cline{2-5}
                  &  $jj\gamma\gamma$ & $\mathtt{MG5\_aMC@NLO}$/$\mathtt{PYTHIA8}$ & $1.40\times 10^4$ & LO &  \\ \cline{2-5}
                  &  $b\bar{b}j\gamma$ & $\mathtt{MG5\_aMC@NLO}$/$\mathtt{PYTHIA8}$ & $2.72 \times 10^5$ & LO &  \\ \cline{2-5}
                  &  $c\bar{c}j\gamma$ & $\mathtt{MG5\_aMC@NLO}$/$\mathtt{PYTHIA8}$ & $9.17 \times 10^5$ & LO &  \\ \cline{2-5}
                  &  $b\bar{b}jj$  & $\mathtt{MG5\_aMC@NLO}$/$\mathtt{PYTHIA8}$ &$3.00 \times 10^8$ & LO &  \\ \cline{2-5}
                  & $Z(\rightarrow b\bar{b})\gamma\gamma$ & $\mathtt{MG5\_aMC@NLO}$/$\mathtt{PYTHIA8}$ & $5.03$ & LO & \\ \hline
\multirow{2}{*}{$t\bar{t}$ and $t\bar{t}\gamma$ BG} & $t\bar{t}$
 & $\mathtt{POWHEG-BOX}$/$\mathtt{PYTHIA8}$  &
$5.30 \times 10^5$ & NNLO &  $\mathtt{CT10}$  \\
 &  &  & & \!\!\!\!\!$+$NNLL &    \\ \cline{2-6}
    ($\geq 1$ lepton)    & $t\bar{t}\gamma$ 
 & $\mathtt{MG5\_aMC@NLO}$/$\mathtt{PYTHIA8}$  &
$1.60 \times 10^3$ & NLO & $\mathtt{CTEQ6L1}$ \\ \hline\hline
\end{tabular}
\end{table}
The Higgs bosons in the signal event samples
are generated on-shell with zero width by
\texttt{POWHEG-BOX-V2}~\cite{Heinrich:2017kxx,Heinrich:2019bkc} with
the damping factor $\mathtt{hdamp}$ set to the default value of 250
to limit the amount of hard radiation. 
This code provides NLO distributions matched to a parton shower 
taking account of the full top-quark mass dependence. 
The variation of the trilinear Higgs coupling, $\lambda_{3H}$, is 
also allowed in this code. 
The Higgs and the top quark masses are set to the default values of
$M_H = 125$ GeV and $m_t = 173$ GeV, respectively,  and
the bottom quark is considered massless. 
The \texttt{MadSpin} code \cite{Artoisenet:2012st} is used after generating a pair of 
Higgs bosons
in order to decay both Higgs bosons into two bottom quarks and two photons. 
For parton showering and hadronization, \texttt{PYTHIA8}~\cite{Sjostrand:2014zea}
is used. Here an appropriate 
setup provided by \texttt{POWHEG-BOX-V2} is used to correctly perform a 
matching of \texttt{POWHEG-BOX-V2} with \texttt{PYTHIA8}.
Finally, fast detector simulation and analysis at the HL-LHC are
performed using \texttt{Delphes3}~\cite{deFavereau:2013fsa}
with the ATLAS template.
The parameters in the template are tuned as in
Ref.~\cite{Chang:2018uwu}.

For generation and simulation of backgrounds, 
we closely follow Ref.~\cite{Chang:2018uwu}, except for the use
of the post-LHC PDF set of {\tt CT14LO}~\cite{Dulat:2015mca} and merged cross sections
for non-resonant backgrounds.
More precisely,
for the two main non-resonant backgrounds of $b\bar b\gamma\gamma$ and $c\bar c\gamma\gamma$,
we use the merged cross sections and distributions
by MLM matching~\cite{Mangano:2006rw,Alwall:2007fs} 
with {\bf xqcut} and $Q_{\rm cut}$ set to 20 GeV and 30 GeV, respectively.
For the remaining non-resonant backgrounds, 
we are using the cross sections and distributions obtained by applying 
the generator-level cuts listed in Eq.~(\ref{eq:genecut}) as 
adopted in Ref.~\cite{atlas_hh17}
which might provide more reliable and conservative estimation of the non-resonant backgrounds
containing light jets~\cite{Chang:2018uwu}. 

The information on the matrix-element generation, parton
showering, and hadronization is summarized in Table~\ref{tab:ParticleList}.
The signal cross section at NNLO order in QCD is calculated according to
\begin{equation}
\sigma^{\rm NNLO} ( \lambda_{3H} )=
K^{\rm NNLO / NLO}_{\rm SM}\, \sigma^{\rm NLO}  (\lambda_{3H}) \;,
\end{equation}
where $\lambda_{3H}$-dependent NLO cross section of
$\sigma^{\rm NLO}(\lambda_{3H})$ is computed by the use of {\tt POWHEG-BOX-V2}
and we take $K^{NNLO/NLO}_{\rm SM}=1.116$~\cite{Grazzini:2018bsd}.
For the cross sections of non-resonant  
and $t\bar t\gamma$ backgrounds,
the following generator-level cuts are applied at parton level
in order to remove the divergence associated with the photons or jets:
\begin{eqnarray}
\label{eq:genecut}
P_{T_j} > 20\  \text{ GeV},\ P_{T_b} > 20\  \text{ GeV},\ P_{T_\gamma} > 25\  \text{ GeV},\ P_{T_l} > 10\  \text{ GeV}, \nonumber\\
 |\eta_j|<5,\ |\eta_\gamma|<2.7,\ |\eta_l|<2.5,\ \Delta R_{jj,ll,\gamma\gamma,\gamma j,jl, \gamma l} > 0.4, \nonumber\\
M_{jj} > 25\ \text{GeV},\ M_{bb} > 45\ \text{GeV},\ 60<M_{\gamma\gamma} < 200\ \text{GeV}.
\end{eqnarray}
Note that, in Table~\ref{tab:ParticleList},
signal and the $ggH(\to\gamma\gamma)$ and
$t\bar t$ backgrounds are generated at NLO and normalized to the
cross sections computed at the accuracy denoted in `Order in QCD'. 
And the
remaining backgrounds are generated at LO and normalized to the
cross sections computed at the accuracy denoted in `Order in QCD'.  

\section{tmva analysis}

\begin{table}[t!]
\caption{Sequence of event pre-selection criteria applied
in this analysis.}
\label{tab:presel}
\begin{center}
  \begin{tabular}{|c|l| }
  \hline
    Sequence &~ Event Pre-Selection Criteria \\
    \hline
    \hline
    1 &~ Di-photon trigger condition,
$\geq $ 2 isolated photons with $P_T > 25$ GeV, $|\eta| < 2.5$
\\
    \hline
    2 &~ $\geq $ 2 isolated photons with $P_T > 30$ GeV,
$|\eta| < 1.37$ or $1.52 < |\eta| <2.37$, 
$\Delta R_{\gamma\gamma,j\gamma} > 0.4$ \\
    \hline
    3 &~ $\geq$ 2 jets identified as b-jets with leading(sub-leading) 
$P_T > 40(30)$ GeV, $|\eta|<2.4$, $\Delta R_{bb} > 0.4$ \\
    \hline
    4 &~ Events are required to contain $\le 5$ jets with
 $P_T >30$ GeV within $|\eta|<2.5$ \\
\hline
    5 &~ No isolated leptons with $P_T > 25$ GeV, $|\eta| <2.5$ \\
    \hline
    \end{tabular}
\end{center}
\end{table}

Before performing a multivariate analysis using 
Toolkit for Multivariate Data Analysis (TMVA)~\cite{TMVA2007}
with {\tt ROOTv6.18}~\cite{ROOT}, 
a sequence of event selections is applied to the signal and background event samples, 
see Table~\ref{tab:presel}.
And then we choose the following eight kinematic variables for TMVA:
\begin{equation}
M_{bb}\,, \ \ P_{T}^{bb}\,, \ \ \Delta R_{bb}\,; \ \
M_{\gamma\gamma}\,, \ \ P_{T}^{\gamma\gamma}\,, \ \ \Delta R_{\gamma\gamma}\,; \ \
M_{\gamma\gamma bb}\,, \ \ \Delta R_{\gamma b}\,.
\end{equation}
We observe that significance can be meaningfully improved by judiciously choosing
the two photons or two $b$ quarks for the above TMVA variables.
For $M_{bb,\gamma\gamma}$ and $P_T^{bb,\gamma\gamma}$, in terms of $P_T$, 
we choose the least energetic two photons or two $b$ quarks 
while the most energetic ones are chosen for $\Delta R_{bb,\gamma\gamma}$ and
$M_{\gamma\gamma bb}$. For $\Delta R_{\gamma b}$, on the other hand, we choose the least
energetic $b$ and the next-to-the-least energetic photon.

\begin{figure}
\centering
\includegraphics[width=2.in]{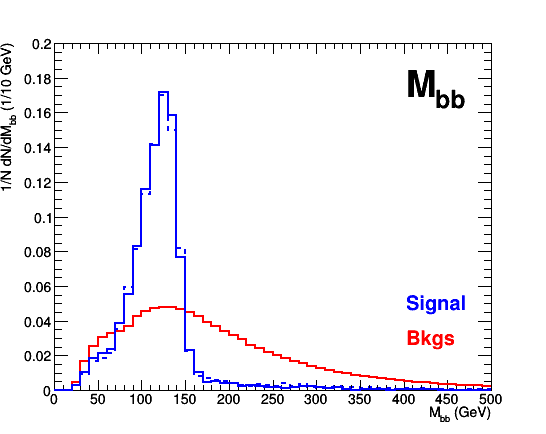}
\includegraphics[width=2.in]{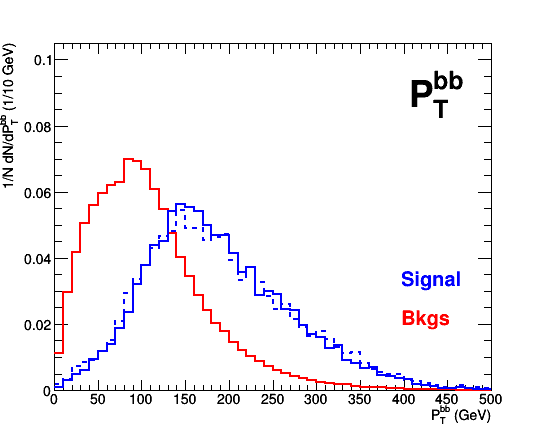}
\includegraphics[width=2.in]{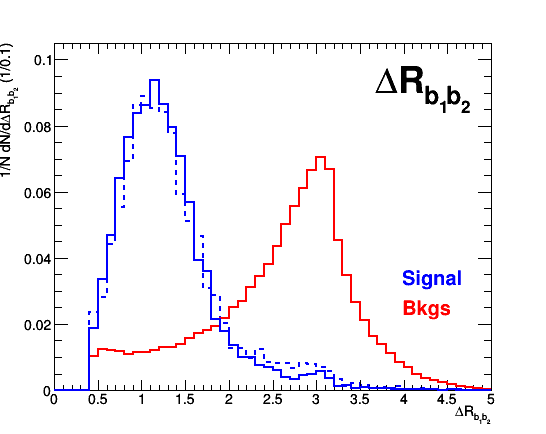}
\includegraphics[width=2.in]{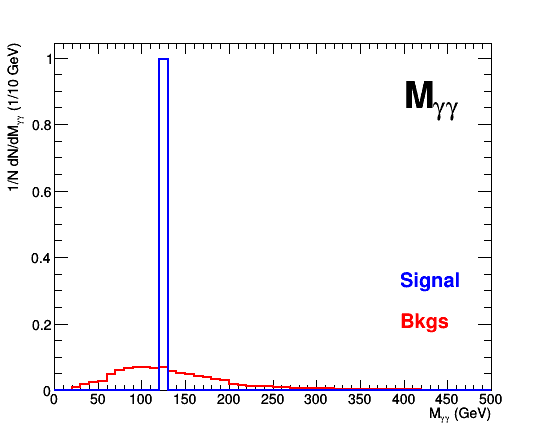}
\includegraphics[width=2.in]{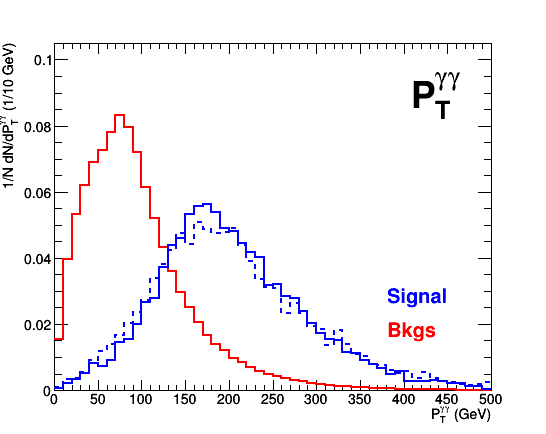}
\includegraphics[width=2.in]{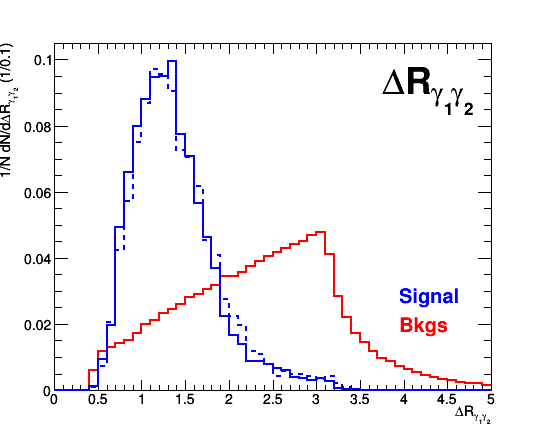}
\includegraphics[width=2.in]{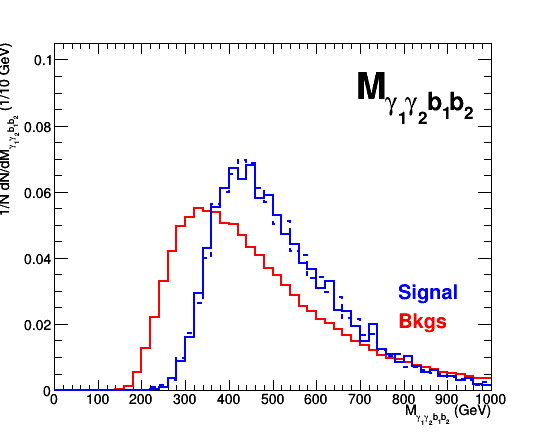}
\includegraphics[width=2.in]{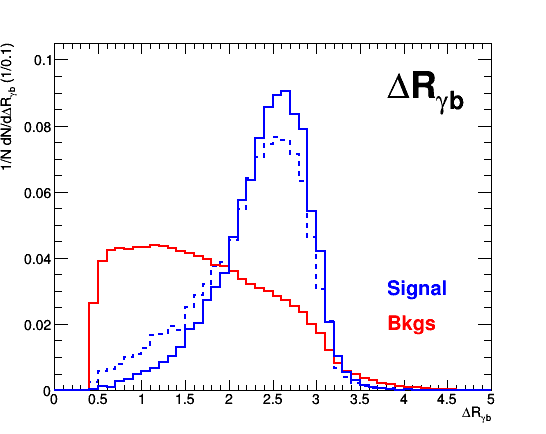}
\caption{Normalized distributions of the eight kinematic variables for TMVA
  for the SM signal with $\lambda_{3H}=1$ (blue) and backgrounds (red)
after applying the event pre-selections cuts in Table~\ref{tab:presel}
\label{fig:TMVA_var}.
For comparisons, the LO signal distributions are also shown in dashed-blue lines.
}
\end{figure}
In Fig.~\ref{fig:TMVA_var}, we show the 
normalized distributions of the eight kinematic variables 
for the SM signal with $\lambda_{3H}=1$ (blue) and backgrounds (red)
after applying the event pre-selection cuts in Table~\ref{tab:presel}.
We observe the broad peak around 125 GeV in the $M_{bb}$ distribution
of the signal
while the peak in the $M_{\gamma\gamma}$ distribution of the signal
is very sharp.
The signal tends to give larger
transverse momenta of $P_T^{bb,\gamma\gamma}$ while it is more populated
in the region of smaller $\Delta R_{bb,\gamma\gamma}$,
implying a strong negative correlation between $P_T$ and $\Delta R$.
Furthermore, the signal has larger $M_{\gamma\gamma bb}$ and its distribution
is peaked around
400 GeV, and $\Delta R_{\gamma b}$ provides another good discriminant observable.

\begin{figure}
\centering
\includegraphics[width=5in]{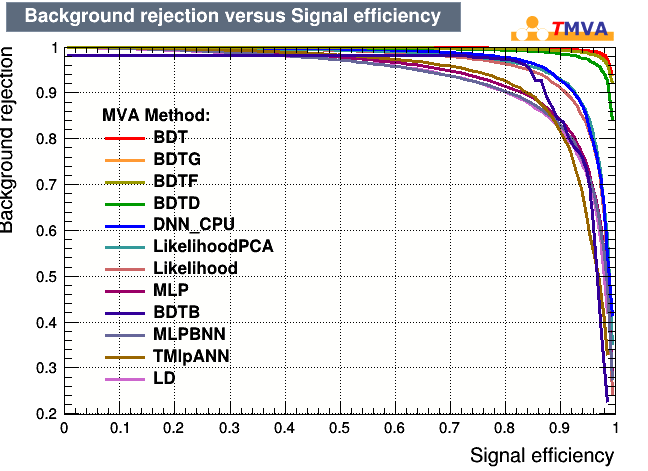}
\caption{
The Receiver Operating Characteristic (ROC) curves
of different multivariate analysis (MVA) methods provided by TMVA.}
\label{fig:BDT_mm}
\end{figure}
First of all, we try various multivariate analysis (MVA) methods provided by 
TMVA with the eight kinematic variables listed above. 
For this we use the default TMVA setup for each method.
The receiver operating characteristic (ROC) curves for various methods are 
shown in Fig.~\ref{fig:BDT_mm}. We find that
the BDT-related methods show higher performance
with better signal efficiency and stronger background rejection.
We choose the best method of BDT for our analysis.

Before presenting the results of our analysis, we describe our setup for BDT briefly here.
For each event sample of signal and backgrounds, we randomly divide it into 
two halves  with a default split seed. 
The first half is used for training and the second one for testing.
For this, we use the following commands:
$$
 \texttt{nTrain$\_$Signal=0:nTrain$\_$Background=0:SplitMode=Random:NormMode=None:!V}\,.
$$
Then in order to improve performance of a trained BDT,
we use 800 trees and node splitting is allowed only
when the number of events in a node is larger than $2.5\%$ of total number of events
of the training sample.
Maximum tree depth is set to 4.
Training is carried out using Adaptive Boost with learning rate $\beta= 0.5$.
One
half of the training sample is randomly chosen at the end of each boosting iteration.
The cut value on the variable in a node is optimized by comparing
the separation index of the parent node and the sum of the indices of 
the two daughter ones. In our work, we choose Gini Index for the separation index.
Finally, the whole range of the variable is equally gridded into 20 cells.
The specific commands used for the performance improvement of BDT training 
are as follows:
\begin{eqnarray}\label{BDT_setup}
 &&\texttt{NTrees=800:MinNodeSize=2.5$\%$:MaxDepth=4:BoostType=AdaBoost:AdaBoostBeta=0.5:}\nonumber \\
 &&\texttt{UseBaggedBoost:BaggedSampleFraction=0.5:SeparationType=GiniIndex:nCuts=20}\,.
\nonumber
\end{eqnarray}

\section{results}
\begin{figure}
\centering
\includegraphics[width=3.12in]{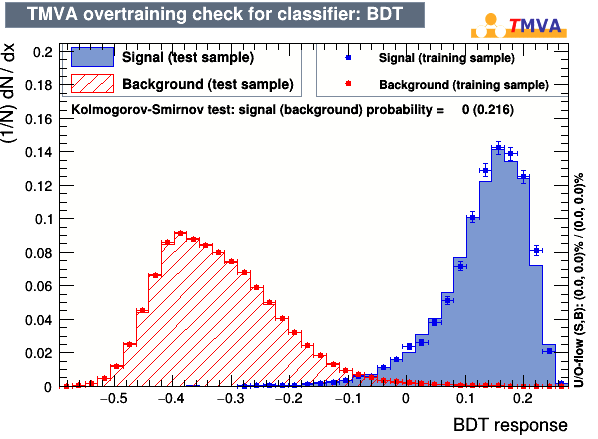}
\includegraphics[width=3.12in]{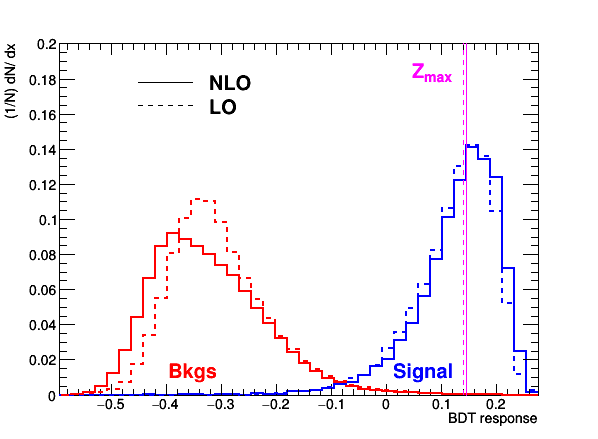}
\caption{
(Left) Normalized SM BDT responses for test (histogram) and training (dots with error bars) 
samples. BDT responses for signal (blue) and background (red) samples mostly populates
in the regions with positive and negative BDT response, respectively.
(Right) Normalized BDT responses for test sample obtained by using NLO (solid)
and LO (dashed) distributions of the eight TMVA input variables.
The vertical lines show the position of 
the optimal cut on the BDT response which maximizes the significance.
}
\label{fig:BDT_over}
\end{figure}
In the left panel of Fig.~\ref{fig:BDT_over}, we show the BDT responses obtained using
BDT trained for $\lambda_{3H}=1$ which is to be called BDT$_{\rm SM}$ shortly.
By validating the BDT distributions for the training sample (dots with error bars)
with those for the test sample (histogram), we check that BDT$_{\rm SM}$ is not overtrained. 
In the right panel of Fig.~\ref{fig:BDT_over}, we compare the BDT responses
for the test sample obtained using NLO (solid)
and LO (dashed) distributions of the eight kinematic variables for TMVA.
We observe that the NLO BDT distributions provide quite better separation between
signal and background.
Incidentally, by the vertical lines, we denote the position of the optimal cut on 
the BDT response which maximizes the significance of $Z$:
\begin{equation}
\label{eq:Z}
 Z = \sqrt{ 2 \cdot \left[ \left( (s+b) \cdot \ln( 1+ s/b) - s \right)
  \right ] }
\end{equation}
where $s$ and $b$ represent the numbers of signal and background events, respectively.

\begin{figure}
\centering
\includegraphics[width=4.2in,height=3.8in]{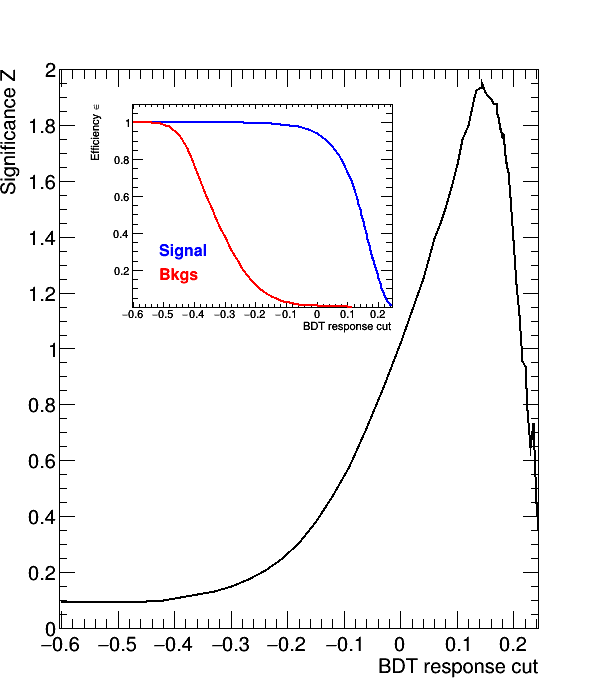}
\caption{
Signal and background efficiencies (inset) and
significance $Z$ as functions of BDT response cut. BDT$_{\rm SM}$ is used.
} \label{fig:effs_BDTL1}
\end{figure}
In Fig.~\ref{fig:effs_BDTL1}, using BDT$_{\rm SM}$,
we show the behavior of signal and background efficiencies
and significance $Z$ according to the variation of the cut value on BDT response.
We observe the significance can reach up to 1.95 when $0.145$ is taken for
the BDT response cut and, at which, the signal and background efficiencies are
$0.50$ and $4.9\times 10^{-4}$, respectively.

\begin{table}[th!]
\caption{
Expected number of signal and background events at the HL-LHC assuming
3000 fb$^{-1}$ using BDT$_{\rm SM}$ with the BDT response cut of $0.145$.
We consider the four representative values of $\lambda_{3H}$ for signal
and the backgrounds are separated into three categories.
For comparisons, we also show the results obtained using the cut-and-count
analysis~\cite{Chang:2018uwu}.
}
\vspace{3mm}
\label{tab:l3h1_opt}
\begin{center}
\begin{tabular}{| l || r | r || r |}
\hline
&\multicolumn{3}{c|}{Expected yields $(3000~ \mathrm{fb}^{-1})$}
\\
\hline
Signal and Backgrounds & ~Pre-Selection & ~BDT$_{\rm SM}$ &  ~Cut-and-Count  \\ \hline
$H(b\,\bar{b})\,H(\gamma\,\gamma)$, $\lambda_{3H} = -5$ &223.22  & 72.73 &90.19 \\
$H(b\,\bar{b})\,H(\gamma\,\gamma)$, $\lambda_{3H} = 0$ & 33.69 & 14.63 & 16.70 \\
$\mathbf{H(b\,\bar{b})\,H(\gamma\,\gamma)}$, $\mathbf{\lambda_{3H}= 1}$ & \textbf{17.77} &
\textbf{8.85} & \textbf{9.63}   \\
$H(b\,\bar{b})\,H(\gamma\,\gamma)$, $\lambda_{3H} = 5$ & 26.37 & 4.04&6.77 \\
\hline
$gg\,H(\gamma\,\gamma)$ & 70.72 &3.94 &7.04  \\
$t\,\bar{t}\,H(\gamma\,\gamma)$ & 157.20  & 3.64&13.14 \\
$Z\,H(\gamma\,\gamma)$ &23.60 & 2.27  &3.60\\
$b\,\bar{b}\,H(\gamma\,\gamma)$ & 2.65 &  0.08 & 0.13\\
\hline
$b\,\bar{b}\,\gamma\,\gamma$ & 4676.36 &3.72 &10.92\\
$c\,\bar{c}\,\gamma\,\gamma$ &  3787.00 &1.20 & 5.41\\
$j\,j\,\gamma\,\gamma$ & 1015.48 &0.39 &2.89 \\
$b\,\bar{b}\,j\,\gamma$ & 10017.91 &0.82 &13.91  \\
$c\,\bar{c}\,j\,\gamma$ & 4679.36& 0.55 &4.78\\
$b\,\bar{b}\,j\,j$ & 2517.71 & 0.05&3.83\\
$Z(b\,\bar{b})\,\gamma\gamma$ & 184.07&  0.32&0.88 \\
\hline
$t\,\bar{t}$~($\geq$ 1 leptons)&  7338.84  & 0.35&5.09 \\
$t\,\bar{t}\,\gamma$~($\geq$ 1 leptons) & 2369.11 &0.62 & 3.69 \\
\hline
Total Background&36839.99& 17.94 & 75.31 \\
\hline\hline
Significance $Z$, $\lambda_{3H} = 1$ && \textbf{1.95} & \textbf{1.09} \\
\hline
\end{tabular}
\end{center}
\end{table}

In Table~\ref{tab:l3h1_opt}, we present
expected number of signal and background events at the HL-LHC assuming
3000 fb$^{-1}$ using BDT$_{\rm SM}$ with the BDT response cut of $0.145$.
We find that the significance is 1.95 with about $9$ signal and $18$
background events for $\lambda_{3H}=1$. Comparing to the results using
the cut-and-count analysis~\cite{Chang:2018uwu}, we find that the number of signal events
decreases by only 10\% while the number of backgrounds by almost 80\%, resulting in 
an increase in significance from $1.09$ to $1.95$.
Note that the composition of backgrounds changes drastically by the use of BDT.
In the cut-and-count analysis, the non-resonant background is about two times 
larger than the single-Higgs associated background. 
While, in the BDT analysis, 
the single-Higgs associated background is larger than the non-resonant one
and $t\bar t$ associated background becomes negligible.
Also, we find that the Higgs boson self-coupling can now be constrained to 
$1.00 < \lambda_{3H} < 6.22$ at 95\% confidence level (CL),
which removes the region of negative $\lambda_{3H}$
in contrast to the results based on the cut-and-count analysis.

Even the significance standing at $1.95$ may not be high enough to make a
precise measurement of the trilinear Higgs self-coupling at the HL-LHC,
we implement a likelihood fitting of $M_{\gamma\gamma bb}$ distribution
to quantify the uncertainty in the determination of $\lambda_{3H}$
and to see how much the two-fold ambiguity in the determination
could be lifted up.

\begin{figure}[th!]
\centering
\includegraphics[width=5.2in]{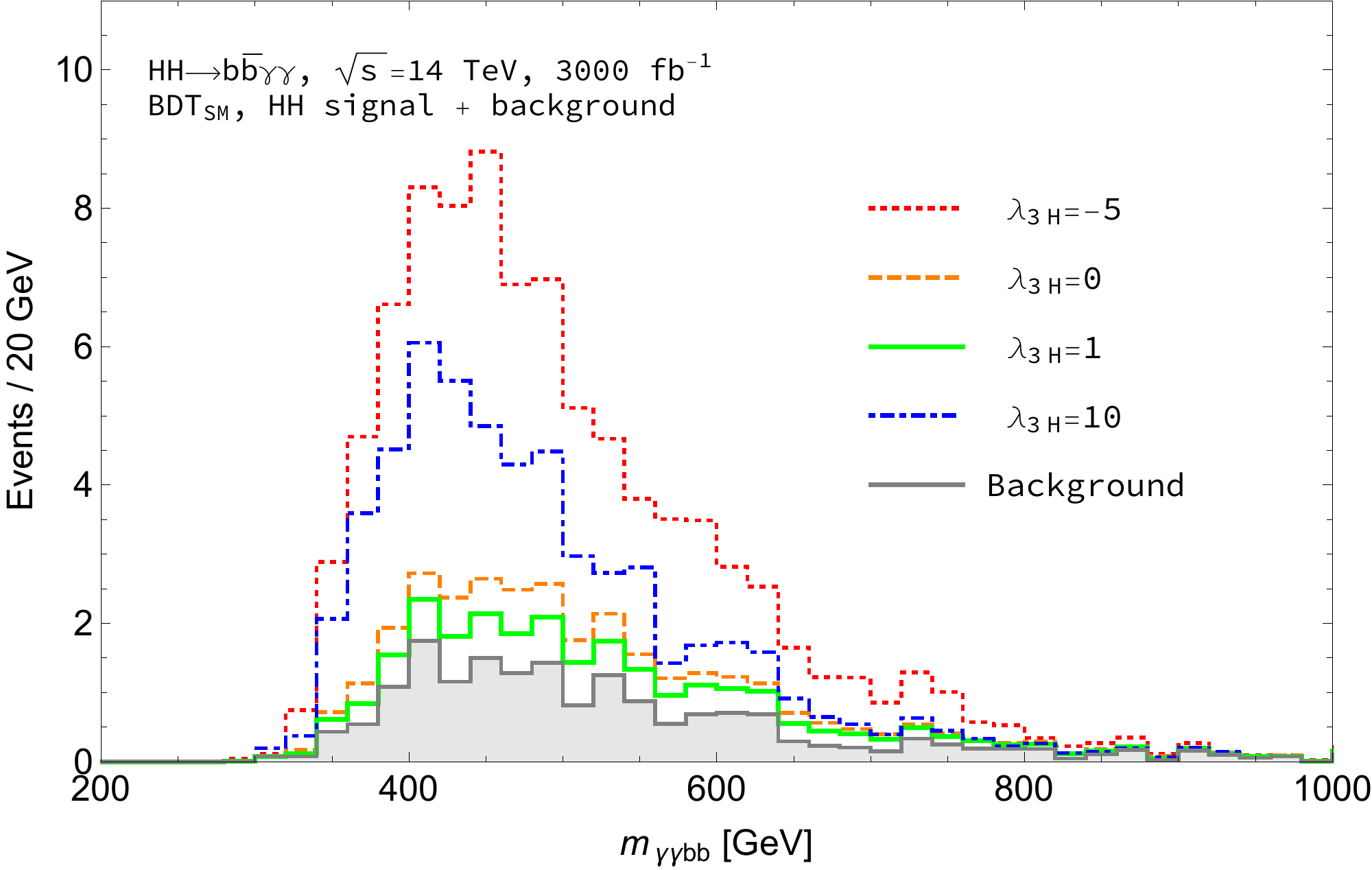}
\includegraphics[width=5.2in]{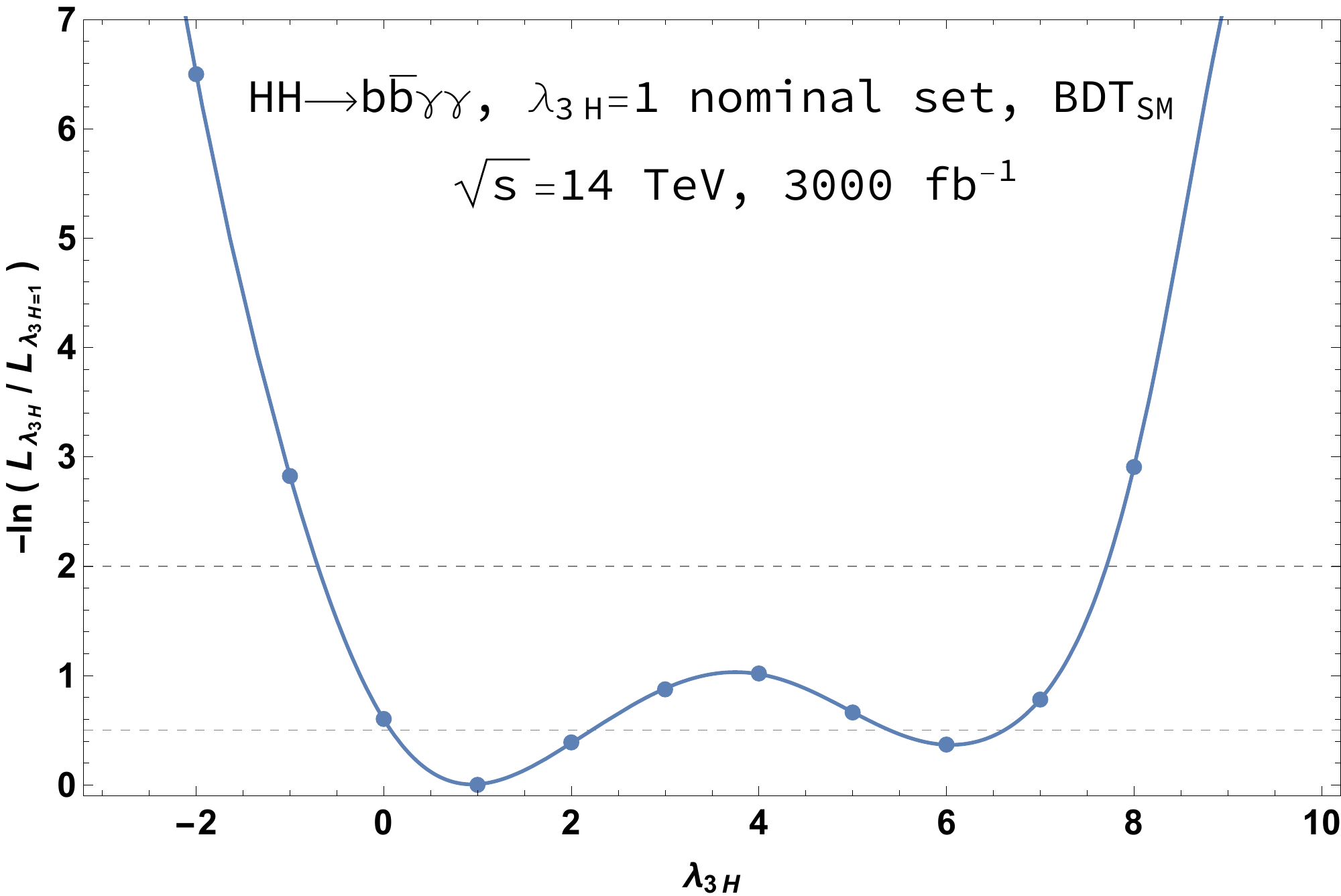}
\caption{ 
(Upper) $M_{\gamma\gamma bb}$ distributions at HL-LHC for the signal and background  
combined, assuming 3000 fb$^{-1}$ and using BDT$_{\rm SM}$ with the BDT response cut of 
$0.145$.
(Lower) The relative log likelihood distribution 
for the nominal value of $\lambda_{3H}=1$.
The dashed line at $0.5\,(2.0)$ indicate the values corresponding to
a $1\sigma\,(2\sigma)$ confidence interval.
}
\label{fig:mhh_BDTL1} 
\end{figure}
In the upper frame of Fig.~\ref{fig:mhh_BDTL1}, we show
$M_{\gamma\gamma bb}$ distributions
for the signal and background combined taking
the representative four values of $\lambda_{3H}$.
The shaded histogram shows the distribution with background only.
In the lower frame of Fig.~\ref{fig:mhh_BDTL1},
we show the log likelihood distribution obtained by 
fitting the signal-plus-background $M_{\gamma\gamma bb}$ distributions
with the nominal value of $\lambda_{3H}=1$.
The solid line shows the result of a polynomial fitting and
we find the $1\sigma$ confidence interval (CI) of
$0.1 < \lambda_{3H} < 2.2\, \cup\,5.4 < \lambda_{3H} < 6.6$.
We observe that the two-fold ambiguity is slightly lifted up
by the amount of $\Delta\left[-\ln(L_{\lambda_{3H}}/L_{\lambda_{3H}=1})\right]
\simeq 0.4$.

\begin{figure}
\centering
\includegraphics[width=5in]{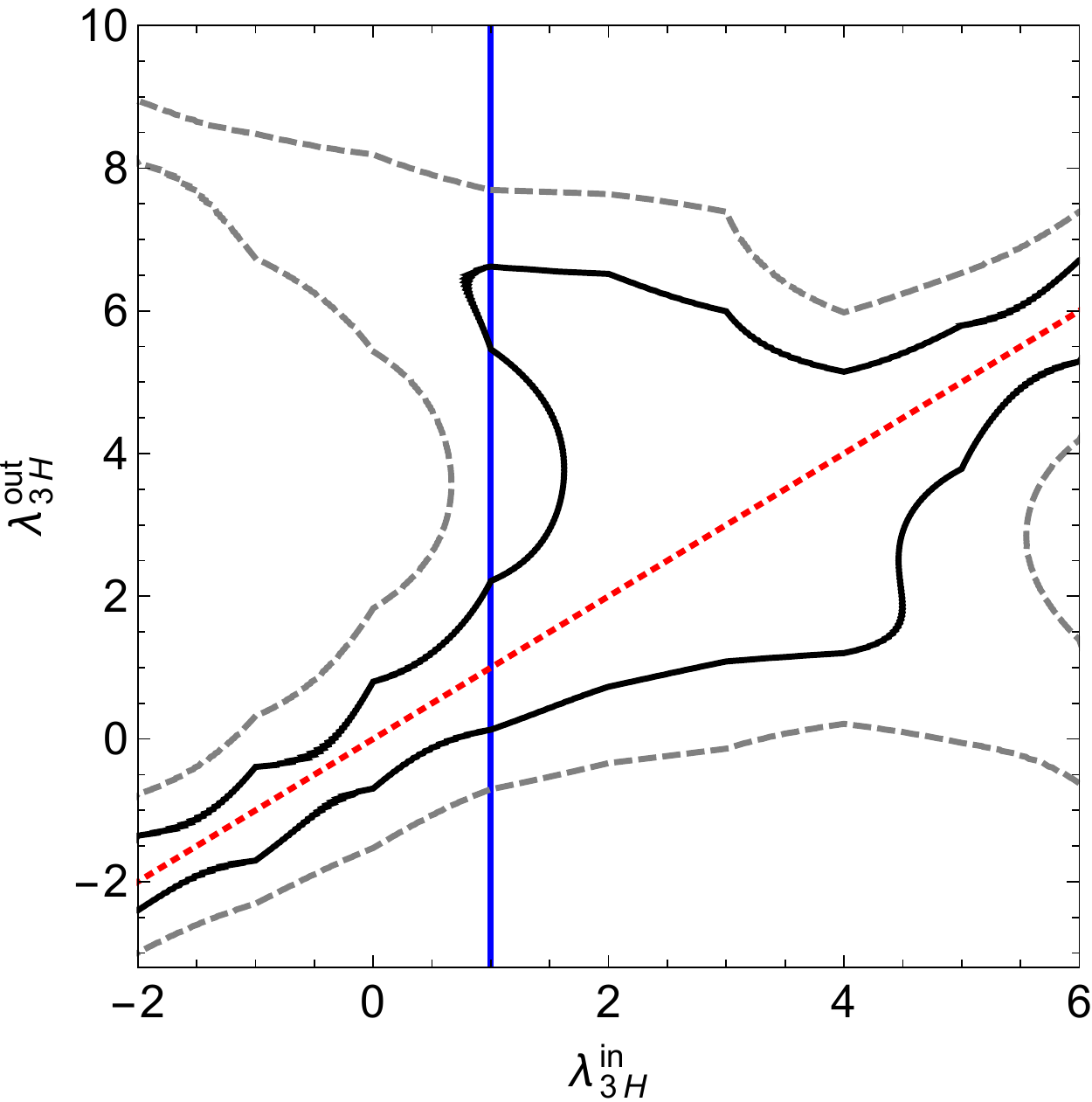}
\caption{
The 1- and 2-$\sigma$ CI regions versus the input  values of
$\lambda_{3H}^{\rm in}$ at the HL-LHC assuming 3 ab$^{-1}$.
The CI regions are obtained by likelihood
fitting of $M_{\gamma\gamma b b}$ distributions 
using BDT$_{\lambda_{3H}}$.
%
The solid  and dashed lines delimit
the 1- and 2-$\sigma$ CI regions, respectively.
And the diagonal red dotted line denotes 
$\lambda_{3H}^{\rm out}=\lambda_{3H}^{\rm in}$
and the vertical blue line at $\lambda_{3H}^{\rm in}=1$ indicates 
the case shown in the lower frame of Fig.~\ref{fig:mhh_BDTL1} 
with BDT$_{\rm SM}$.
}
\label{fig:Contour_BDTLX}
\end{figure}
    So far we have used the
    BDT trained for $\lambda_{3H}=1$ or BDT$_{\rm SM}$
independently of the input value of $\lambda_{3H}$.
Without knowing the value of $\lambda_{3H}$ a priori, 
it would be more desirable to use separate BDTs trained for 
specific values of $\lambda_{3H}$, which we wish to call
BDT$_{\lambda_{3H}}$ in short for further references.
In Fig.~\ref{fig:Contour_BDTLX}, we show the contour plot showing the
$1\sigma$ and $2\sigma$ CI regions obtained by likelihood
fitting of $M_{\gamma\gamma b b}$ distributions.
For each value of $\lambda_{3H}^{\rm in}$, 
we use the corresponding BDT$_{\lambda_{3H}}$  together with
$\lambda_{3H}=\lambda_{3H}^{\rm in}$ nominal set and
the BDT response cut is set to maximize significance.
Using BDT$_{\lambda_{3H}}$, the 95\% CL region is narrowed into
$1.01 < \lambda_{3H} < 5.42$ at 95\% CL. 
Compared to that obtained using BDT$_{\rm SM}$,
we observe the noticeable
changes of $Z$ for $\lambda_{3H}\gsim 4$.
And we also find there exists a bulk region 
of $0.5 \lsim \lambda_{3H} \lsim 4.5$ in which it is hard for one to
pin down the trilinear coupling, see 
the 1-$\sigma$ error region delimited by solid lines in Fig.~\ref{fig:Contour_BDTLX}.

\begin{table}[th!]
\caption{
NLO and LO results obtained using BDT$_{\rm SM}$ and comparison with
the recent ATLAS result~\cite{atlas_pub-2018-053}.
$\mathrm{Z}_{\mathrm{max}}$, $s|_{_{\mathrm{Z}_{\mathrm{max}}}}$, and
$b|_{_{\mathrm{Z}_{\mathrm{max}}}}$ denote the significance,
the number of signal events, and the number of background events, respectively,
obtained after applying the BDT cut which maximizes the significance.
Also compared are the 95$\%$ CL and 1$\sigma$ CI ranges of $\lambda_{3H}$.
}
\vspace{3mm}
\label{tab:Comparison_Table}
\begin{center}
\begin{tabular}{| c || r | r || r |}
\hline
& \multicolumn{2}{c||}{BDT$_{\rm SM}$} & ATLAS$\_$2018 \cite{atlas_pub-2018-053} \\
\hline
& MLM $\oplus$ NLO$_{\rm dist.}$
& MLM $\oplus$ LO$_{\rm dist.}$
& Ref.~\cite{atlas_hh17} $\oplus$ LO$_{\rm dist.}$ \\
\hline
\hline
$\mathrm{Z}_{\mathrm{max}}$ & 1.95 & 1.79 & 2.1 \\
\hline
$s|_{_{\mathrm{Z}_{\mathrm{max}}}}$  & 8.85 & 8.61 & 6.46\\
\hline
$b|_{_{\mathrm{Z}_{\mathrm{max}}}}$  &  17.94 & 20.44 & 6.8 \\
\hline
\hline
95$\%$ CL & $(1.00, 6.22)$ & $(0.87, 6.55)$ & $(1.3, 6.2)$ \\
\hline
1$\sigma$ CI &  $(0.1, 2.2)\, \cup\, (5.4, 6.6)$ & $(0.1, 2.3)\, \cup\, (5.3, 6.7)$ & $(-0.1,
2.4)$ \\
\hline
\end{tabular}
\end{center}
\end{table}
In Fig.~\ref{fig:BDT_over}, we show that using the NLO distributions of 
signal for TMVA may lead to the better results. For a quantitative comparison,
we use the LO distributions for TMVA and
find that the significance can reach up to 1.79 with about $9$ signal and $20$
background events.
The details of the NLO and LO results based on BDT$_{\rm SM}$ are presented in
Table~\ref{tab:Comparison_Table} where we also make comparison with
the recent ATLAS result without systematic uncertainties~\cite{atlas_pub-2018-053} 
in which the LO distributions of signal and the generator-level cuts
Eq.~(\ref{eq:genecut}) adopted in Ref.~\cite{atlas_hh17}
are used for TMVA.

Lastly, we consider the impacts of the TMVA random seed used to
divide each event sample of signal and backgrounds into 
the training and test samples and the Monte Carlo (MC) random seed for the signal 
event samples. And we check that the fluctuation of the significance 
due to the changes of random seeds is negligible.
%

%

\section{conclusions}
Higgs-pair production is the most useful avenue to the understanding of the EWSB
sector. We have studied in great details, with the help of machine learning, the
sensitivity of measuring the trilinear Higgs self-coupling $\lambda_{3H}$
that one can expect at the HL-LHC with 3000 fb$^{-1}$. With TMVA one can improve
upon the signal-to-background significance over the traditional cut-based analysis.
In this work, we have shown that the significance is improved by about 80\% and
found a narrower range of $\lambda_{3H}$ below the sensitivity.
With  BDTs trained for each value of $\lambda_{3H}$, we found the bulk region
down to $0.5 \alt \lambda_{3H} \alt 4.5$ in which one cannot pin down
the trilinear coupling.

\section*{Acknowledgment}
We thank 
Chih-Ting Lu for initial participation.
%
%
This work was supported by the National Research Foundation of Korea
(NRF) grant No. NRF- 2016R1E1A1A01943297.
K.C. was supported by the
MoST of Taiwan under grant number MOST-105-2112-M-007-028-MY3
and 107-2112-M-007-029-MY3.
J.P. was supported by the NRF grant No. NRF-2018R1D1A1B07051126.

\newpage


\end{document}